\documentclass[aps,prb,twocolumn,secnumarabic,groupedaddress]{revtex4}

\usepackage{graphicx,epsfig}
\usepackage{amssymb,amsmath}
\usepackage{color}
\usepackage[usenames,dvipsnames]{xcolor}
\usepackage[colorlinks=true,linkcolor=Red,citecolor=Green,linktoc=page]{hyperref}
\begin{document}

\title{Quantum $H$-theorem and irreversibility in quantum mechanics}

\author{
	G.\,B.\,Lesovik,$^1$
	I.\,A.\,Sadovskyy,$^2$
	A.\,V.\,Lebedev,$^3$
	M.\,V.\,Suslov$^4$ and
	V.\,M.\,Vinokur$^2$
}

\affiliation{
	$^1$L.D.~Landau Institute for Theoretical Physics RAS,
	Akad. Semenova av., 1-A, Chernogolovka, 142432, Moscow Region, Russia
}

\affiliation{
	$^2$Materials Science Division, Argonne National Laboratory,
	9700 S. Cass Avenue, Argonne, Illinois 60637, USA
}

\affiliation{
	$^3$Theoretische Physik, Wolfgang-Pauli-Strasse 27, ETH Z\"urich, CH-8093 Z\"urich, Switzerland
}
	
\affiliation{
	$^4$Moscow Institute of Physics and Technology,
	Institutskii per. 9, Dolgoprudny, 141700, Moscow District, Russia
}

\date{\today}

\begin{abstract}
We investigate temporal evolution of von Neumann's entropy in exemplary quantum mechanical systems and show that it grows in systems evolving with incrementally increasing decoherence during scattering processes. We demonstrate that the origin of irreversibility lies in complexity of preparing time-reversed quantum states due to entanglement and in partitioning of the wave function of the evolving system.
\end{abstract}

\maketitle

\section{Introduction}

\begin{figure}[b]
	\begin{center}
		\includegraphics[width=8.2cm]{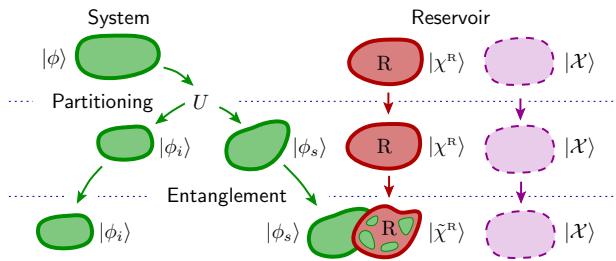}
	\end{center} \vspace{-4mm}
	\caption{
		Partitioning and entanglement.
		We divide every interaction event between the system (green) and the reservoir
		(red and magenta) into two steps, unitary evolution and entanglement with the reservoir.
		Unitary evolution is typically accompanied by an increase in the number of the
		components of the wave function, partitioning, and does not change the system entropy.
		Entanglement additionally promoted by preceding partitioning leads to entropy growth.
		The part of the reservoir, R (red), interacting with the system has component
		$|\chi^{\rm\scriptscriptstyle R}\rangle$ , and $|{\mathcal X}\rangle$ is the
		component corresponding to the noninteracting part of the reservoir (magenta).
	}
	\label{fig:two_processes}
\end{figure}

In the 1870-s, Ludwig Boltzmann published his celebrated kinetic equation and the $H$-theorem~\cite{Boltzmann:1872,Boltzmann:1896} that gave the statistical foundation of the second law of thermodynamics~\cite{Lebowitz:1999}. In 1877, to better explain his $H$-theorem, Boltzmann introduced the statistical entropy~\cite{Boltzmann:1877} $S = k_{\scriptscriptstyle\rm B} \log W$, where $k_{\scriptscriptstyle\rm B}$ is the Boltzmann constant, and $W$ is the {\it Wahrscheinlichkeit}, the number of possible microstates corresponding to the macroscopic state of a given thermodynamic system. He offered a proof that $dS/d\tau \geqslant 0$, based on the {\it Stosszahlansatz}, the molecular chaos hypothesis, having stated that velocities of colliding particles are uncorrelated and independent of position. John von Neumann asserted that irreversibility arises within quantum mechanics without invoking molecular chaos. His proof~\cite{Neumann:1929} of non-decreasing entropy defined as $S = - k_{\scriptscriptstyle\rm B} {\rm tr}\{{\hat\rho}\log{\hat\rho}\}$, where ${\hat\rho}$ is the density matrix of the quantum system in question, still used a procedure of macroscopic measurement, leaving, therefore, a question whether the entropy growth can be established within the pure quantum approach open. Recent developments related the origin of irreversibility to the entanglement phenomenon~\cite{Gemmer:2004,Gemmer:2001,Popesku:2006}. It was shown that, in an isolated system, $S$ is constant~\cite{Nielsen:2011}, while letting the quantum system interact with an environment leads to the increase in the entropy via entanglement~\cite{Gemmer:2001,Gemmer:2004,Winter:2006,Linden:2009}. However, the mechanisms by which either monotonic or non-monotonic temporal behaviour of entropy would arise, and the conditions under which the quantum $H$-theorem holds, remains unclear. Here we prove the quantum $H$-theorem, which contains von Neumann's formulation as a limiting case, and show that non-decreasing entropy evolution emerges due to interaction of the system with the environment accompanied by wave function {\it partitioning} and its {\it entanglement} with the reservoir. Furthermore, we demonstrate that irreversibility arises solely within unitary quantum mechanics due to reversal procedure complexity.

\section{Quantum $H$-theorem}

Following Boltzmann's insight attributing an origin of the entropy growth to particles interactions,
we describe a temporal evolution of any system belonging in the realm of quantum physics as a sequence of distinct scattering events, which, in general, include interactions of the quantum system with a reservoir. Notice that one can split every scattering event into two stages, see Fig.~\ref{fig:two_processes}. Stage~A, where unitary evolution of the system wave function $|\phi\rangle$, accompanied, in general, by partitioning, i.e., by increasing the number of the wave function components. Importantly, during stage~A system's entropy does not change. Stage~B comprises interaction of the system with the reservoir accompanied by entanglement of the system wave function with the reservoir's degrees of freedom. Now we formulate quantum $H$-theorem:

{\it If (i)~there exists a basis in which the diagonal elements of the density matrix remain unchanged in the course of the interaction with the reservoir degrees of freedom during stage~B and (ii)~the system interacts with each degree of freedom of the reservoir only once, the system entropy is nondecreasing.}

Importantly, the condition (i) of quantum $H$-theorem is not only sufficient but the necessary one. Indeed, if the condition (i) is not satisfied, the system is not isolated from the reservoir even from the classical viewpoint and no general statements about entropy are possible. At the same time, violation of the condition (ii) in general may, although not necessarily does, lead to decreasing entropy during the system evolution, see examples below.

The $H$-theorem can be cast into more formal terms by noticing that at the stage~B the interaction with the reservoir, preserving the classical distribution function, can be described as the following transformation of the wave function $|\psi\rangle$ of a grand system comprising the evolving quantum system and the reservoir:
\begin{equation}
	|\psi\rangle = \sum\nolimits_i c_i |\phi_i\rangle |\chi_i\rangle
	\quad \Rightarrow \quad
	|{\tilde\psi}\rangle = \sum\nolimits_i c_i |\phi_i\rangle |{\tilde\chi}_i\rangle,
	\label{eq:psi}
\end{equation}
where the set $\{|\phi_i\rangle\}$ is the system orthonormal basis so that $\langle\phi_i|\phi_j\rangle = \delta_{ij}$, $\{|\chi_i\rangle\}$ is the set of the wave functions describing the reservoir, $\langle\chi_i|\chi_i\rangle = 1$, and $\sum\nolimits_i |c_i|^2 = 1$; the states $|\chi_i\rangle$ are, in general, not orthogonal. Modified reservoir wave functions are $|{\tilde\chi}_i\rangle$, $\langle{\tilde\chi_i} | {\tilde\chi}_i\rangle = 1$. Let us introduce the {\it decoherence matrix} $\hat\beta$ with components
\begin{equation}
	\beta_{ij}\equiv 
	\langle{\tilde\chi}_i|{\tilde\chi}_j\rangle / \langle\chi_i|\chi_j\rangle
	\label{eq:beta}
\end{equation}
and correspondent transformation $\hat\rho \to \hat{\tilde{\rho}} \equiv \hat\beta \circ \hat\rho$, where `$\circ$' denotes Schur element-by-element matrix multiplication, $\rho_{ij} \to \tilde\rho_{ij} \equiv \beta_{ij} \rho_{ij}$. The diagonal elements of the decoherence matrix are equal to unity, which exactly expresses the fact that classical distribution function given by diagonal elements of the density matrix $P_i = \rho_{ii}$ is not affected by this transformation (distribution itself changes during partitioning).

Initial disentanglement from the reservoir means that~$|\chi_i\rangle$ can be presented in a form $|\chi_i\rangle=|\chi_i^{\rm\scriptscriptstyle R}\rangle|\mathcal{X}\rangle$, where~$|\chi_i^{\rm\scriptscriptstyle R}\rangle$ describe the part of the reservoir that was interacting in the past and got entangled with the system and $|\mathcal{X}\rangle$ refers to the part of the reservoir which was not interacting so far and will get entangled in the course of upcoming interaction. Therefore, $\beta_{ij} = \langle{\tilde\chi}_i^{\rm\scriptscriptstyle R}|{\tilde\chi}_j^{\rm\scriptscriptstyle R}\rangle$ and $\hat\beta_{ij}$ is a Gramian (positive semidefinite) matrix. Note, that in more general case the part of the reservoir $|\chi\rangle$ that is disentangled from the system might be in a mixed state (entangled with some other system). Then quantum $H$-theorem follows from the Lemma:

{\it Transformation $\hat\rho \to \hat{\tilde{\rho}} \equiv \hat\beta \circ \hat\rho$ with Gramian decoherence matrix leads to $S(\hat{\tilde{\rho}})\geqslant S(\hat{\rho})$.}

Here we present a simple straightforward proof of the Lemma for the infinite-dimensional density matrix. Let density matrix be parametrized by some continuous variables $s$ and $s^\prime$: $\langle s|\hat\rho|s^\prime\rangle = \rho[s,s^\prime]$. Let us introduce an auxiliary quantum system (reservoir) $F$, whose initial state is $|F\rangle = \sum_n f(q_n) |q_n\rangle$ with $\langle q_n| q_m \rangle = \delta_{nm}$ and $\sum_n |f(q_n)|^2 = 1$. Let our original system be in the state $|\Phi\rangle = \int ds\, a(s)\, |s\rangle$ with $\langle s| s^\prime\rangle = \delta(s-s^\prime)$. Assuming that the interaction leads to the following unitary transformation $\hat{U} (|s\rangle |q_n\rangle) = e^{i\phi(q_n,s)} |s\rangle |q_n\rangle$, the two systems get entangled,
\begin{equation}
	|\Phi\rangle |F\rangle \to 
	\int ds \sum\nolimits_n a(s) f(q_n)\, e^{i\phi(q_n,s)} |s\rangle 
	|q_n\rangle.
	\label{eq:entstate}
\end{equation}
Tracing out the behavior of the auxiliary system one generates the following transformation of the initial density matrix,
\begin{align}
	{\tilde\rho}[s,s^\prime]
	& = \sum\nolimits_n |f(q_n)|^2\,
		e^{i\phi(q_n,s) - i\phi(q_n,s^\prime)}\,\rho[s,s^\prime] \nonumber \\
	& = \beta(s,s^\prime) \rho[s,s^\prime],
	\label{eq:ctr1}
\end{align}
where $\beta(s,s^\prime) = \langle F(s)|F(s^\prime)\rangle$ is the Gramian matrix for the transformed reservoir states, $|F(s)\rangle = \sum_n f(q_n) e^{i\phi(q_n,s)}|q_n\rangle$. In this case:

{\it If $\rho[s,s^\prime]$ and ${\tilde\rho}[s,s^\prime]$ are given by transformation~\eqref{eq:ctr1} with $\sum_n |f(q_n)|^2 = 1$, then $S({\tilde\rho}[s,s^\prime]) \geqslant S(\rho[s,s^\prime])$.}

{\it Proof:} Using convexity of von Neumann entropy one finds
\begin{eqnarray}
	&&S\Bigl( \sum\nolimits_n |f(q_n)|^2 e^{i\phi(q_n,s)} \rho[s,s^\prime] e^{-i\phi(q_n,s^\prime)} \Bigr)
	\nonumber\\
	&&\qquad \geqslant\sum\nolimits_n |f(q_n)|^2 S\Bigl( e^{i\phi(q_n,s)} \rho[s,s^\prime] e^{-i\phi(q_n,s^\prime)} \Bigr)
	\nonumber\\
	&&\qquad = \sum\nolimits_n |f(q_n)|^2 S\bigl( \rho[s,s^\prime] \bigr) = S\bigl( \rho[s,s^\prime] \bigr),
\end{eqnarray}
where in the last line we have used the property that a unitary transformation of the density matrix $\rho[s,s^\prime] \to e^{i\phi(q_n,s)} \rho[s,s^\prime] e^{-i\phi(q_n,s^\prime)}$ does not change the von Neumann entropy.$\quad\square$

In this proof we used the additional phase prefactors only, see Eq.~(\ref{eq:entstate}). It can be generalized for arbitrary Gramian matrix by utilizing additional amplitude modifications. Note that extending Uhlmann's theory onto the infinite-dimensional case following Wehrl~\cite{Wehrl:1978}, one generalizes the above consideration to the continuous $q$ limit and replace $\sum_n \to \int dq$. The similar Lemma was formulated in the recent works on quantum information theory, where it was shown that when transmitting information through the unital channel, entropy is nondecreasing~\cite{Holevo:2010}.

\subsection{Examples of physical systems, where $H$-theorem holds}

Let us illustrate the above general findings by exemplary manifestations of the $H$-theorem. We start with an important example of electrons in disordered conductors in the case where the time of the energy relaxation $\tau_\varepsilon$ well exceeds the dephasing time $\tau_\varphi$~\cite{Altshuler:1982}. This time scales separation naturally provides a possibility for splitting the electron interaction with the reservoir (environment) into two stages introduced above. In order to demonstrate how the conditions for validity of the $H$-theorem are fulfilled in this case we model a single event of electron scattering by its interaction with a high potential barrier. Here, we consider the case of the interaction of an electron with the reservoir having zero temperature. In this case the interaction is reduced to emission of photons by scattered electrons ({\it Bremsstrahlung} or braking irradiation). For the sake of simplicity, we consider a one-dimensional electron wave-packet $\phi(x,\tau)$ propagating towards an infinite barrier located at $x = 0$. After the scattering the electron density matrix $\rho=\phi(x,\tau) \phi^*(x^\prime,\tau)$ transforms into
\begin{eqnarray}
	\tilde\rho(x,x^\prime,\tau) =
	\rho \exp(-\Phi(x-x^\prime))
	\label{eq:photon}
\end{eqnarray}
with
\begin{equation}
	\Phi(x) = \frac{8\alpha_0}{3\pi}\,\frac{v^2}{c^2}\,
	\int\nolimits_0^\Omega d\omega\, \frac{1-\exp({\rm i}\omega x / v_{\rm\scriptscriptstyle F})}{\omega},
	\label{eq:phi}
\end{equation}
where $\alpha_0$ is a fine structure constant, $c$ is the speed of light, and $\Omega$ is an ultra-violet cutoff, corresponding to the maximal possible frequency of the emitted photon. Equation (\ref{eq:phi}) is derived by linearizing electron's energy-dispersion relation near Fermi velocity $v_{\rm\scriptscriptstyle F}$. The above transformation of the density matrix is a continuous analog of Eq.~(\ref{eq:ctr1}) with $\phi(q_n,s) \to s q$. Since the emitted photons run away they do not affect the subsequent dynamics of the electron. Considering further scattering events during the diffusive motion the electron density matrix~$\tilde{\rho}$ will experience the similar transformation for the each next scattering. These subsequent transformations lead to the monotonic increase of the electron entropy during its motion~\cite{Lebedev:2014}. Note, that although the diagonal elements of the density matrix in coordinate representation do not change, see Eq.~(\ref{eq:photon}), they do change in either momentum or energy representation. This corresponds to the transfer of the small part $\delta\epsilon \propto \alpha_0 (v_{\rm\scriptscriptstyle F}/c)^2 E_{\rm\scriptscriptstyle K}$ of the electron's kinetic energy $E_{\rm\scriptscriptstyle K}$ to the electromagnetic field. Another natural example is qubits where the decay time of the diagonal elements, $T_1$, well exceeds the decay time of the off-diagonal elements, $T_2$~\cite{Burkard:2004}. Then at times $\tau\sim T_2$, the effect of the interaction with the reservoir is that while the off-diagonal elements of the qubit density matrix nearly vanish, whereas the diagonal ones remain practically intact. Thus the qubit density matrix obeys the conditions for the $H$-theorem.

\section{Arrow-of-time}

\begin{figure}
	\begin{center}
		\includegraphics[width=6.0cm]{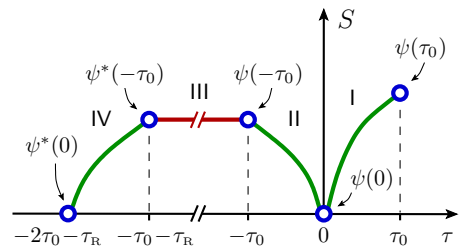}
	\end{center} \vspace{-4mm}
	\caption{
		Global entropy symmetry diagram.
		The system is prepared in the non-entangled state at $\tau=0$. Forward 
		evolution occurs with increasing entropy (temporal segment~I).
		Backward evolution, segment II, occurs with also with increasing entropy 
		which thus is larger at $\tau = -\tau_0$ than entropy at $\tau=0$. 
		During the time-interval III, we create a non-trivial entangled state via full 
		conjugation $\psi(-\tau_0) \to \psi^*(-\tau_0)$. Then, the backward temporal 
		evolution leads to the entropy decrease during the interval IV.
	}
	\label{fig:symmetry}
\end{figure}

\begin{figure*}[!t]
	\begin{center}
		\includegraphics[width=14.0cm]{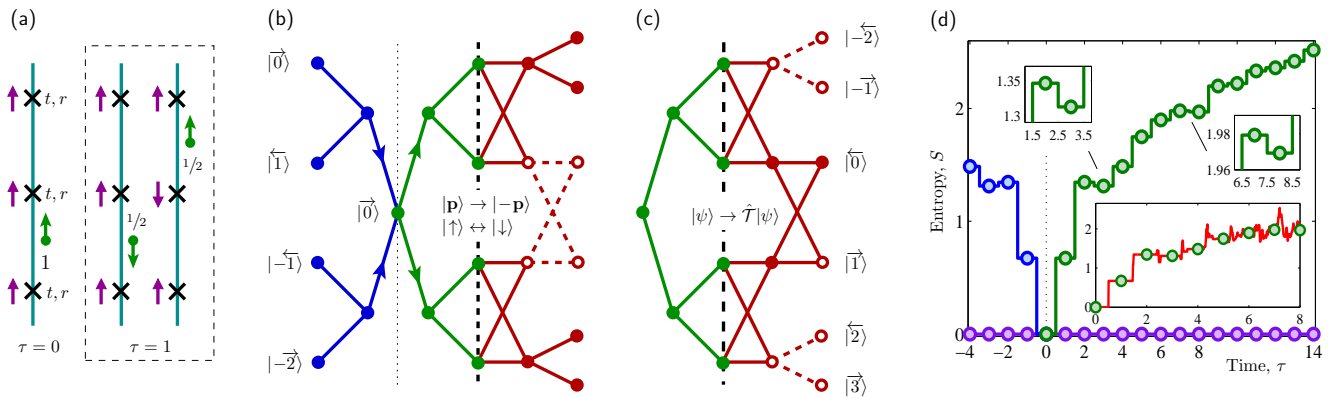}
	\end{center} \vspace{-8mm}
	\caption{
		Forward and backward evolution of one-dimensional system.
		(a)~Model system setup. A particle (green) propagating through a regular
		array of scatterers associated with spins (magenta) and characterized by transmission
		and reflection amplitudes $t$ and $r$, respectively, which are the same for each
		scatterer, $|t|^2 = 0.5$. A spin flips if the particle passes the corresponding scatterer
		and remains intact if it is reflected. At every time step spins are
		replaced by a new set. Two possibilities at first step are shown.
		(b)~Incomplete conjugation: after two steps we inverse velocities and spins
		and the system evolves two steps `backward' (red). The circles are the components of
		the wave function, the filled circles are non-zero, the empty circles that result from the
		destructive interference are equal to zero for chosen $t =\exp({\rm i}\pi/4)/\sqrt{2}$, and
		$r =\exp(-{\rm i}\pi/4)/\sqrt{2}$, but can be non-zero for other values of $t$'s and $r$'s.
		Each circle corresponds to some quantum state, we indicate it for initial and final states
		omitting numerical coefficients and spin part. The lines emerging from the circle to the
		right denote the coherent splitting of the wave packet at the scatterer into the transmitted
		and reflected ones; the lines that enter circles from the left denote the incoming wave
		packets parts. Dashed lines correspond to the zero amplitude process at chosen $t$
		and $r$. After two steps forward all basis vectors $|\chi_i\rangle|\phi_i\rangle$ are
		reversed by the time reversal operator $\hat{\mathcal T} =\exp({\rm i}\pi/2) \sigma_y
		\hat{\mathcal K}$ (in our case this leads to flipping all spins and reversal of the direction
		of particle motion in the central region between the scattering centers), where
		$\hat{\mathcal K}$ is complex conjugation operator and $\sigma_y$ is the Pauli matrix.
		The amplitudes $c_i$ of these basis vectors are not modified. Conjugation of basis
		vectors is not sufficient for reversing the evolution: some components of the wave
		function survive, and non-trivial entanglement between electron and spins forms.
		Additionally, the symmetric in time evolution of the wave function of the grand system
		is shown in blue.
		(c)~Reversibility. Complex conjugation of all amplitudes in addition to that of the
		basis vectors implies conjugation of the wave function of the system. Then the evolution
		is reversible.
		(d)~Entropy of the 1D system (green) as a function of time $\tau$, for $|t|^2=0.4$
		monotonically increases `on average,' but locally may drop. The circles show the entropy
		between scattering/dephasing events. Both forward (green) and backward (blue) temporal
		evolution processes lead to the increase in entropy. The spin-sets are not replaced after
		the scattering event. At $\tau=0$ the particle and spins are not entangled and the particle
		is localized between the two adjacent scatterers. The entropy's behavior is symmetric in
		time. If the reservoir is absent, entropy remains zero (magenta). The inset represents
		entropy evolution in the 1D system with randomness in the positions of scatterers.
	}
	\label{fig:1d_evolution}
\end{figure*}

Having established the origin of the quantum $H$-theorem as the increasing entanglement of the system's wave function with the reservoir and the related decoherence, one can ask, where exactly the time reversal symmetry, which is seemingly built into the Schr\"odinger equation, gets lost. In order to answer this question and demonstrate global time symmetry, let us consider a grand system (comprising the quantum subsystem and the reservoir) described by the wave function~$\psi(\tau)$ and governed by the time-independent Hamiltonian~$\hat{\mathcal H}$ (we assume here $\hat{\mathcal H} = \hat{\mathcal H}^*$). Let us suppose that for any non-entangled product state of the grand system, the system entropy grows during the forward-time evolution, i.e., for any product state $\psi(\tau) = \phi\otimes\chi$ and $\tau' > \tau$ the condition $S(\psi(\tau')) > S(\psi(\tau))$ holds, where $\psi(\tau') = \hat{U}(\tau' - \tau) \psi(\tau)$ and $\hat{U}(\tau) = \exp(-{\rm i}\hat{\mathcal H}\tau)$. We denote $S(\psi) \equiv S({\rm tr}_{\rm\scriptscriptstyle R} \{|\psi\rangle\langle\psi|\})$, where ${\rm tr}_{\rm\scriptscriptstyle R}$ is a trace over reservoir degrees of freedom. Let the grand system be in some specifically prepared product state $\psi(0)$ at time $\tau = 0$. Then at preceding times $\tau < 0$ the evolution that started at $\tau = -\tau_0 < 0$ and continued till $\tau = 0$ and is governed by the time-independent Hamiltonian (excluding the process of state preparation) occurs with {\it decreasing} entropy. Shown in Figs.~\ref{fig:symmetry}, \ref{fig:1d_evolution}(b), and \ref{fig:1d_evolution}(d) are examples of the future-past symmetric evolutions, see the related discussion in Appx.~\ref{sec:entropy_symmetry} and Ref.~\onlinecite{Lesovik:2013}. Indeed, $\psi(-\tau_0) =\exp({\rm i}\hat{\mathcal H}\tau_0) \psi(0)$ and thus $\psi^*(-\tau_0) = \exp(-{\rm i}\hat{\mathcal H}\tau_0) \psi^*(0)$. The latter corresponds to the forward temporal evolution of the state $\psi^*(0)$ and thus, according to our assumption, to $S(\psi^*(-\tau_0)) = S(\psi(-\tau_0)) > S(\psi(0))$. Now we see that the loss of time symmetry should occur at the moment of state preparation at $\tau = 0$, where the Hamiltonian ceases to be time-independent. Had it been possible to prepare the grand system in the state $\psi(-\tau_0)$, the system entropy would have decreasing in the course of evolution from $\tau = -\tau_0$ till $\tau = 0$, see Fig.~\ref{fig:symmetry}. However, such a preparation can hardly be realized in practice since it implies the complex conjugation of the naturally entangled state $\hat{U}(\tau_0)\psi^*(0)$. The complex conjugation of an unknown quantum state cannot be realized by a unitary operation~\cite{Lesovik:2013} (except for some special cases, e.g., spin echo phenomenon~\cite{Hahn:1950} and qubit dynamic decoupling~\cite{Viola:1999}). Then one needs to prepare the conjugated state by hand, provided the full knowledge of the many-particle entangled wave-function of the grand system is available. Such a reversal procedure requires an access to the reservoir's degrees of freedom, which are usually inaccessible; this is commonly referred to as dissipation. Since the preparation of the state evolving with decreasing entropy on demand for a thermodynamically isolated macroscopic system is an enormously complex (if realistic at all) problem, the probability of spontaneous emergence of such a state is vanishingly low. Note, that these states evolving with decreasing entropy violate the conditions of the formulated above $H$-theorem, since preparation of these states implies that the system was interacting with the same degrees of freedom of the reservoir in the past. In reality, the most probable scenario of fluctuations with entropy decrease is realized by the energy exchange with a reservoir if one observes the system for times longer then~$\tau_\epsilon$.

Finally, we would like to stress here that the systems we discuss do not belong in the class covered by the Poincar\'e recurrence theorem~\cite{Poincare:1890,Bocchieri:1957}. In particular, we consider systems possessing continuous or discrete but infinitely degenerate spectra.

\subsection{Demonstration of the reversal complexity}

\begin{figure}
	\begin{center}
		\includegraphics[width=8.6cm]{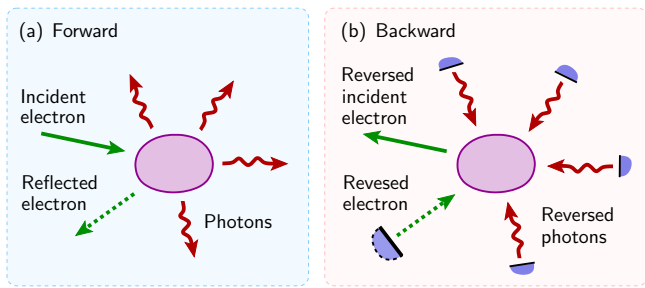}
	\end{center} \vspace{-4mm}
	\caption{
		Forward and backward scattering process.
		(a)~An electron, representing a quantum system, reflects on the scatterer. 
		Reflection is accompanied by braking radiation and electron becomes entangled 
		with emitted photons, representing the reservoir. The photon energy is assumed 
		to be low, i.e., no energy and momentum transfer occurs. This process is described 
		by Gramian decoherence matrix and, therefore, is accompanied by entropy increasing.
		(b)~If one can reverse the electron and photon wave functions after reflection 
		(which are already entangled) the system will absorb photons and evolute to the 
		reversed initial electron wave function. The entropy decreases in the reversion process.
	}
	\label{fig:forward_backward}
\end{figure}

Next, we demonstrate the complexity of the reversal problem on a few simple physical models. The first example is the braking radiation due to scattering of an electron on a potential barrier, see Fig.~\ref{fig:forward_backward}. In order to reverse the scattered state, one needs to reverse both the outgoing photon waves and the scattered electron. The practical reversal includes placing a set of phase conjugation mirrors~\cite{Leith:1966,Boyd:2008} around the scatterer in order to convert the diverging wave into the converging one. We have estimated the complexity of the reversal of the electronic wave packet, see Appx.~\ref{sec:wave_packet}, and found the polynomial reversal complexity $\propto \tau^n$ for 1D and spherically symmetric 3D cases. The practical reversal of the photonic waves with linear spectrum so far has been realized with the aid of nonlinear optic effects~\cite{Andreev:2002} only.

In the next example we demonstrate emergence of irreversibility and the realization of the $H$-theorem for the particle interacting with scatterers and associated spins in one-dimensional system, see Fig.~\ref{fig:1d_evolution}(a). Figures~\ref{fig:1d_evolution}(b) and \ref{fig:1d_evolution}(c) compare the scenario that returns the system to the original state after the second step, and the scenario which does not. Namely, the former case implements the complete time reversal procedure, while the latter one comprises the velocities and angular momenta reversal only. According to this scenario the entropy grows on average both forward and backward in time [Fig.~\ref{fig:1d_evolution}(d)], but at some steps, where disentanglement occurs, can drop (small insets). The entropy decrease occurs precisely because of the repeated interaction with the reservoir. Entropy evolution in the system with random spin/scatterer positions, shown in the larger inset in Fig.~\ref{fig:1d_evolution}(d), does not exhibit essential deviations from evolution in the regular array. The details of calculations are given in Appx.~\ref{sec:1d}. Another example of a system where evolution accompanied by disentanglement and thus by decreasing entropy takes place is offered by a single atom initially entangled with a few photons and is described in Ref.~\onlinecite{Haroshe:2008}.

\begin{figure}
	\begin{center}
		\includegraphics[width=8.6cm]{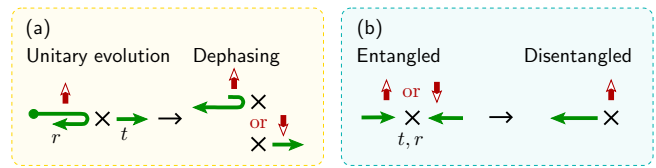}
	\end{center} \vspace{-4mm}
	\caption{
		Violation of the $H$-theorem conditions during disentanglement.
		(a)~An 1D electron, representing a quantum system, can be either transmitted 
		or reflected (with amplitudes~$t=1/\sqrt{2}$ and~$r={\rm i}/\sqrt{2}$ respectively) 
		upon interaction with the scatterer. The scatterer is assigned by a spin so that 
		the spin flips if the electron is transmitted, and remains intact if electron is reflected. 
		After interaction the electron election and spin wave functions become entangled 
		and entropy grows.
		(b)~After reversion of the entangled electron and spin wave functions the system 
		evaluates to the initial state. This reversed evolution occurs with entropy drop 
		to the initial value.
	}
	\label{fig:entanglement_disentanglement}
\end{figure}

In order to illustrate how the conditions of the $H$-theorem get violated during the
time-reversal procedure, let us consider a single interaction event of the scattered electron
with the spin reservoir, see Fig.~\ref{fig:entanglement_disentanglement}. The forward evolution implies the following transformation of the electron density matrix:
\begin{equation*}
	\left[\begin{array}{cc}
		1/2& tr^*\\
		rt^*&1/2
	\end{array}\right] \to
	\left[ \begin{array}{cc}
		1/2 & 0 \\
		0 & 1/2
	\end{array}\right]\!,
\end{equation*}
corresponding to ${\hat\beta} = {\rm diag}\{1,1\}$. Here we set $t=1/\sqrt{2}$ and $r={\rm i}/\sqrt{2}$. The reversal procedure and the subsequent backward time evolution, means that the off-diagonal elements of matrix $\hat\beta$ are divergent and, therefore, $\hat\beta$ is not a Gramian matrix, which violates the condition of Lemma that ensures the $H$-theorem.

\section{Conclusion}
In conclusion, we formulated and proved the quantum $H$-theorem stating that the quantum systems whose temporal evolution is accompanied by increasing their entanglement with the reservoir and thus growing decoherence evolve with non-decreasing entropy. In mathematical terms the sufficient condition for the quantum $H$-theorem to hold can be coined as the requirement for decoherence matrix be Gramian. We demonstrated that the general origin of irreversibility in quantum systems is complexity of actual preparing of the time reversed state.

\subsection*{Acknowledgments}
We are delighted to thank Yu.\,M.\,Galperin, G.\,Blatter, J.\ Fr\"ohlich, G.\,M.\,Graf, R.\,Renner, L.\,B.\,Ioffe, and M.\,McBreen for illuminating discussions and the critical reading of the manuscript. The work was supported by the RFBR Grant No. 11-02-00744-a (G.B.L.), the Pauli Center for Theoretical Studies at ETH Z\"urich (G.B.L.), the U.S. Department of Energy, Office of Science, Materials Sciences and Engineering Division under the Contract No. DE-AC02-06CH11357 (I.A.S. and V.M.V.), and the Swiss National Foundation through the NCCR QSIT (A.V.L.).

\appendix

\section{Symmetry of entropy in time
\label{sec:entropy_symmetry}}
Let us examine what happens to the system in the past at $\tau<0$ if at $\tau=0$ it is not entangled with the reservoir and gets entangled in the course of evolution at $\tau>0$, see Figs.~\ref{fig:symmetry} and \ref{fig:1d_evolution}. We consider, as an example, a particle in three-dimensional space propagating through the random potential and emitting photons. Let the particle wave function be
\begin{equation*}
	\phi({\bf r}) = \int \frac{d{\bf k}}{(2\pi)^3} \exp({\rm i}{\bf kr}) f({\bf k}),
\end{equation*}
where $f({\bf k})$ is a real number function. The coordinate reversal ${\bf r} \to -{\bf r}$ gives rise to the complex conjugation $\phi({\bf r}) \to\phi^*({\bf r})$. If the potential is invariant {\it on average} with respect to ${\bf r} \to -{\bf r}$ transformation then the temporal evolution is symmetric with respect to the time reversal. Since the particle still sees the same potential, the direct evolution of the state $\psi^*$ is the same as that of $\psi$. Therefore $\psi^*(-\tau)=\psi(\tau)$, and $S(\tau)=S(-\tau)$.

To better understand the time symmetry, let us consider the special situation where symmetry in the entropy behavior vanishes. Let quenched disorder occupy a half space. A particle that incidents from the empty half space would scatter and emit photons and thus increase entropy because of the entanglement with photons. Reversing the wave functions of the particle and photons will cause the decrease in entropy till the particle returns into the empty half-space, and then the entropy would remain constant.

\section{Wave packet reversion complexity
\label{sec:wave_packet}}

In this section we discuss the reversal complexity of the wave function in one-dimensional (1D) and three-dimensional (3D) cases.

\subsection{One-dimensional case
\label{sec:wave_packet_1d}}

The wave packet, $\psi(x)$, is defined as
\begin{equation}
	\psi(x) = \int \frac{dk}{2\pi} \exp({\rm i}kx) f(k),
	\label{eq:wp_Fourier}
\end{equation}
where $f(k)$ is the Fourier transform of the packet wave function. At large $\tau$ the wave function $\psi(x,\tau) = \exp(-{\rm i}\mathcal{\hat H}\tau)\psi(x)$ assumes the form
\begin{equation}
	\psi(x,\tau) =
	\int \frac{dk}{2\pi} f(k) \exp({\rm i}kx) \exp\Bigl(-{\rm i} \frac{\hbar k^2}{2m}\tau \Bigr).
	\label{eq:wp_spreading}
\end{equation}
Evaluating the integral at $\tau \to \infty$ one finds\cite{Reed:1975}
\begin{equation}
	\psi(x,\tau) \to
	\frac{f(xm/\hbar\tau)}{\sqrt{2\pi{\rm i}\hbar\tau/m}}
	\exp\Bigl({\rm i} \frac{m x^2}{2\hbar\tau} \Bigr),
	\label{eq:wp_inf_t}
\end{equation}
demonstrating the spreading of the wave packet. Let us fix the time $\tau$, and expand the phase around arbitrary $x=x_0$, setting the smooth function $f(xm/\hbar\tau) \approx f(x_0m/\hbar\tau)$:
\begin{equation*}
	\frac{{\rm i} m x^2}{2\hbar\tau} =
	\frac{{\rm i} m x_0^2}{2\hbar\tau} +
	\frac{{\rm i} m x_0}{\hbar\tau}\Delta x +
	\frac{{\rm i} m}{2\hbar\tau} \Delta x^2,
\end{equation*}
where $\Delta x = x - x_0$.

\begin{figure}[b]
	\begin{center}
		\includegraphics[width=6.4cm]{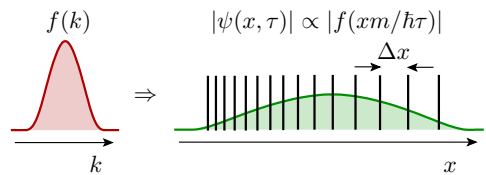}
	\end{center} \vspace{-4mm}
	\caption{
		The array of mirrors demonstrating reversibility of the evolution
		of the wave packet in 1D.
	}
	\label{fig:wave_packet_1d_reversibility}
\end{figure}

After complex conjugation the wave packet gets narrower and after time $\tau$ collapses into an initial one. Let us estimate the accuracy with which we can perform this complex conjugation procedure by switching on $N$ sequential infinite walls (mirrors) located at positions $\{x_n\}$, $n = 1,2, \ldots, N$. After reflecting a part of the wave packet lying between the $n$-th and $(n-1)$-th mirror the $n$-th mirror is switched off. Let us analyze a single reflection event. Setting $x_0 = x_n$ at the initial moment of reflection, we notice that the sign of $\Delta x_n = x - x_n$ changes in Eq.~(\ref{eq:wp_Fourier}). In other words, $\psi(x_n+\Delta x_n,t)$ transforms to $\exp({\rm i}\theta) \psi(x_n-\Delta x_n,t)$, where $\theta$ is some phase depending on mirror and its position. Therefore, the linear part $\exp({\rm i}mx_0\Delta x/\hbar\tau)$ can be exactly reverted. Naturally, one can conjugate the constant factor of the wave function, $\exp({\rm i}mx_0^2/2\hbar\tau)$. But, the factor containing quadratic in $\Delta x$ phase, $\exp({\rm i}m\Delta x^2/2\hbar\tau)$, cannot be reverted by such a `qubit' operation because the quadratic term in the phase does not change the sign. Therefore, in order to implement the reversal procedure, one has to ensure that this term is small enough and can be disregarded. To estimate an error brought by this term, we consider the difference between two arbitrary wave functions $\psi_1$ and $\psi_2$ with the same amplitude, but having the small difference in phase, $\alpha(x)$, $|\alpha(x)| \ll 1$.
The difference (the norm) between these wave functions is
\begin{align*}
	||\psi_1 - \psi_2||^2 & = 2 - 2 \int dx \, |\psi_1(x)|^2 \cos\alpha(x) \\
	& = \int dx \, |\psi_1(x)|^2 \alpha^2(x)
	\leqslant \alpha^2,
\end{align*}
where $\alpha = {\rm max}_x\{|\alpha(x)|\}$. One says then that $\psi_1$ is the same as $\psi_2$ with the accuracy $\varepsilon$ if $\alpha \leqslant \varepsilon$.

Coming back to the quadratic term in the wave function one can say that if one wants to revert the wave packet with the accuracy $\varepsilon$ one should take intervals of the width $\Delta x$ such that $\Delta x^2 m/2\hbar\tau \leqslant \varepsilon$. So the interval width is $\Delta x \propto \sqrt{\varepsilon \tau}$. Since the width of the wave packet is proportional to the evolution time~$\tau$, the number of mirrors (elementary operations) is $N \propto \sqrt{\tau/\varepsilon}$.

The $k$-vectors (and velocities) are different for different~$x$, thus, the mirrors should not be equidistant, see Fig.~\ref{fig:wave_packet_1d_reversibility}. The time of reversing wave packet backward $\propto \sqrt{\tau}$ increases the procedure complexity, but does not change previous estimation dramatically. One can reduce the time necessary for the reversal procedure via placing the additional $\sim \sqrt{\tau}$ mirrors in between of the each pair of mirrors of the original set. Then the total number of mirrors is $\propto \tau$ (as in Ref.~\onlinecite{Lesovik:2013}), and the reversal time becomes the $\tau$-independent constant. Summarizing, the crude estimate shows the power law for wave packet reversal complexity. This justifies our statement about the reversal of the wave-packet evolution in 1D.

Let us discuss the possibility of the more straightforward reversal of the wave packet. Namely, instead of switching on the reflection when the packet passes the scatterer, we rearrange the mirror before the wave packet hits it. The mirror is such that the phase of the reflection amplitude~$\chi = 2E\tau/\hbar - 2kx - 2{\rm arg} f(k)$, $r = \exp({\rm i}\chi)$, strongly depends on the magnitude of the wave vector $k$. We need such a special mirror in order to implement a passive version of so-called phase conjugation mirror\cite{Leith:1966}. The dwelling time at the mirror $\sim \hbar\partial \chi/\partial E$ is of the order of the evolution time~$\tau$ itself. Thus, while passing the mirror, the wave packet would acquire an additional phase $\sim -(\hbar k^2/2m)\tau$. Therefore this phase is to be absorbed into an original phase of the wave packet, and, in its turn, the reflection should be consistently modified giving rise to the additional accumulation of the phase while passing the mirror, resulting in the indefinite growth of the reversal time. Thus even a crude estimate shows that the wave packet reversal with the aid of the passive mirror looks problematic if possible at all.

\subsection{Three-dimensional case
\label{sec:wave_packet_3d}}

\begin{figure}[tb]
	\begin{center}
		\includegraphics[width=4.2cm]{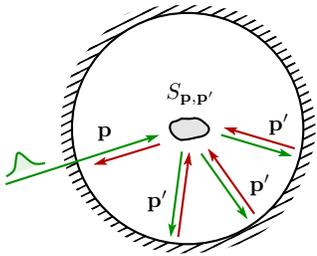}
	\end{center} \vspace{-4mm}
	\caption{
		Reversibility of the wave packet in 3D.
	}
	\label{fig:wave_packet_3d_reversibility}
\end{figure}

Let us evaluate complexity of the reversal of a wave packet interacting with a single scattering center in 3D. The Schr\"odinger equation describing the scattering is
\begin{equation*}
	- \frac{\hbar^2}{2m} \Delta \psi + U({\bf r}) \psi = E \psi,
\end{equation*}
where $U({\bf r})$ is the scattering potential. It is convenient to use a scattering theory relating the amplitudes of the incident $|\bf p\rangle$ and the scattered $|\bf p'\rangle$ waves, where ${\bf p}$ and ${\bf p'}$ are their respective momenta, via the scattering matrix $S_{{\bf p},{\bf p}'}$. Let us choose a coordinate origin at the scattering center, then the scattering of the wave packet with the momenta about $\bf p$ into the state with the momenta about~${\bf p}'$ is represented as
\begin{equation*}
	|{\bf p}\rangle\to\int d{\bf p'}S_{{\bf p},{\bf p'}}|{\bf p'} \rangle,
\end{equation*}
see Fig.~\ref{fig:wave_packet_3d_reversibility}.

After scattering, a diverging spherical wave-front with the radius increasing due to propagation of the packets away from the scatterer, and with the increasing width due to the spreading of the packets, forms. Upon the reversal procedure the wave packet re-assembles itself at the scatterer into the original packet but moving in the opposite direction, provided the scattering matrix is symmetric, $S_{{\bf p},{\bf p'}}=S_{{\bf -p'},{\bf -p}}$. In the Born approximation and far from the scatterer the scattering potential can be viewed as spherically symmetric, thus the problem reduces to the one-dimensional one. Therefore the reversal procedure in 3D retains the polynomial complexity. Yet, in order to complete the process, one has, in addition, to reverse the part of the packet that passes through without scattering, which further complicates the reversal.

\section{Infinite one-dimensional walk
\label{sec:1d}}

\begin{figure}[b]
	\includegraphics[width=8.0cm]{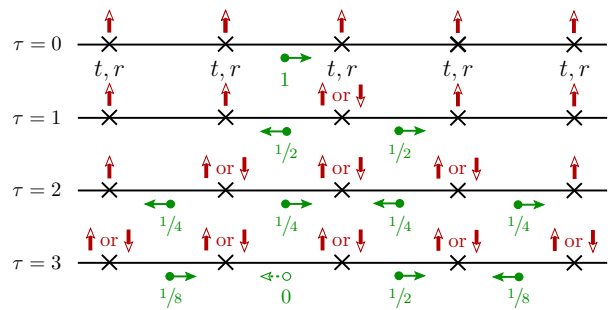}
	\caption{
		Evolution of the infinite 1D system.
	}
	\label{fig:1d_setup_evolution}
\end{figure}

An infinite 1D setup, see Fig.~\ref{fig:1d_setup_evolution}, where a point-like wave packet propagates along an array of scattering centres and the associated spins. The point-like wave packet starts the motion from the position in between the scatterers and moves to the right. We assume, for simplicity, that the spectrum is linear and the wave packet does not spread.

The set of spins represents a reservoir. The entanglement process occurs in the following way: the wave packet that passes the scatterer flips the spin, while the reflected one leaves the spin intact. We further let the energy required for the spin to flip be infinitesimal, so neither the energy transfer nor the momentum exchange occur\cite{Levitov:1996}. The only relevant process is the phase change ensuring the entanglement. Set of spin is not replaced after every scattering event. This implies that the spin system accumulates the memory about the particle evolution. The absence of the walls allows for the particle to propagate indefinitely and interact with the infinite number of spins.

In Sec.~\ref{sec:1d_equidistant} we consider the equidistant scatterers/spins in an infinite 1D model and the discuss the corresponding temporal evolution of the von Neumann entropy. In Sec.~\ref{sec:1d_noiseAmp} we add the random component to transmission/reflection amplitudes. In Sec.~\ref{sec:1d_noise} we include the randomness in scatterers/spin positions. We show that the exponential growth in the number of the wave function components and, thus, the growth of complexity of the time-reversal procedure arises solely from the randomness of the distribution of scattering centres, even in the absence of a spin-related reservoir.

\begin{figure}
	\includegraphics[width=8.5cm]{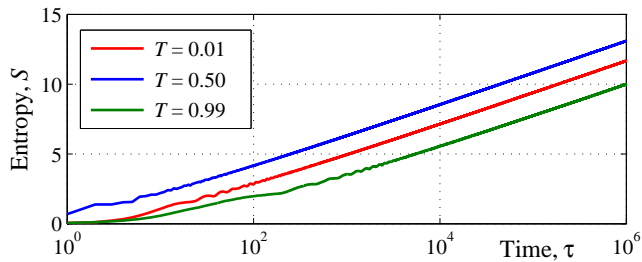}
	\caption{Von Neumann entropy growth in infinite regular 1D system, logarithmic scale.}
	\label{fig:1d_log}
\end{figure}

\subsection{Spin-flip process
\label{sec:1d_spin_flip}}

Let us consider the spin-flip process in more detail. The passing wave packet induces the magnetic field $ {\bf B}(\tau) = [0, B_y(\tau), 0]$ acting on the spin. Let us assume that the field acts during some time while the wave packet dwells in the spin's vicinity. Then the Hamiltonian controlling the evolution is
\begin{equation*}
	\hat{\mathcal H}_{\rm int}(\tau)
	= -\gamma (\hbar/2) {\bf\sigma}{\bf B}(\tau)
	= -\gamma (\hbar/2) \sigma_y B_y(\tau)
\end{equation*}
and the initial state is transformed by the evolution operator
\begin{align*}
	{\hat U} & = \exp\biggl(-\frac{\rm i}{\hbar} \int \hat{\mathcal H}_{\rm int}(\tau) d\tau \biggr)
	= \exp\biggl(\frac{{\rm i} \gamma}{2} \sigma_y \int B_y(\tau) d\tau\biggr) \\
	& = \cos\biggl(\frac{\gamma}{2} \sigma_y \int B_y(\tau) d\tau\biggr)
		+ {\rm i} \sigma_y \sin\biggl(\frac{\gamma}{2} \sigma_y \int B_y(\tau) d\tau\biggr).
\end{align*}
Here $\gamma$ is the gyromagnetic ratio and $\sigma = [\sigma_x, \sigma_y, \sigma_z]$ are Pauli matrices. Then the angle of the spin turn is governed by the magnitude of the interaction time. In particular, the choice $(\gamma/2) \int B_y(\tau) d\tau = \pi/2$ flips the spin. Then the operator
\begin{equation*}
	{\hat U} = {\rm i} \sigma_y =
	\left[\begin{array}{cc}
		0 & 1 \\
		-1 & 0
	\end{array}\right]
\end{equation*}
implements transformations ${\hat U}|\!\uparrow\rangle = |\!\downarrow\rangle$ and ${\hat U}|\!\downarrow\rangle = |\!\uparrow\rangle$, where
\begin{equation*}
	|\!\uparrow\rangle =
	\left[\begin{array}{c}
		1 \\ 0
	\end{array}\right]\!, \qquad
	|\!\downarrow\rangle =
	\left[\begin{array}{c}
		0 \\ 1
	\end{array}\right]\!.
\end{equation*}

\subsection{Equidistant scatterers/spins
\label{sec:1d_equidistant}}

Let the set of scatterers/spins be a regular array. The first few steps of the evolution are shown in Fig.~\ref{fig:1d_setup_evolution}. Initially (at $\tau=0$) the particle is located exactly between two scatterers ($i=0$) and moves to the right. The scattering processes at scatterers occur simultaneously. Therefore we discretize the time~$\tau$ and consider the system at integer time values, while assigning scattering events to the semi-integer times. At the first step ($\tau=1$) the reflected part of the wave packet is located at the same position and is moving to the left (with the reflection probability $R$ and amplitude $r = {\rm i}R^{1/2}$), whereas the transmitted part is located to the right of the initial position and is moving to the right (with the transmission probability $T=1-R=|t|^2$ and amplitude $t = T^{1/2}$). At next steps, we obtain the set of probabilities $P_i^\rightarrow(\tau)$ ($P_i^\leftarrow(\tau)$) for observing the particle at sequential positions labeled by subscript $i$ and moving to the right (to the left). The probability of finding the particle at the position $i$ is $P_i(\tau) = P_i^\rightarrow(\tau) + P_i^\leftarrow(\tau)$. Note that due to the chosen initial conditions, $P_i^\leftarrow(\tau)=0$ if $i+\tau$ is even, $P_i^\rightarrow(\tau)=0$ if $i+\tau$ is odd. Therefore, $P_i(\tau)$ is defined only by the part of wave packet moving either to the right or to the left but not by their superposition. The particle and spin states get entangled, and by measuring the spin states one can determine the particle state, and/or, vice versa, by observing the position of the particle one can recover the spin subsystem state. At the same time the system does not record the history of the particle movements.

\begin{figure}
	\includegraphics[width=8.5cm]{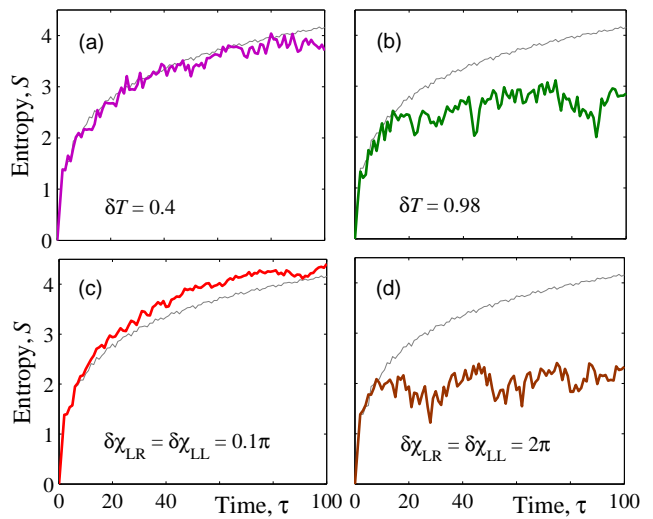}
	\caption{
		Entropy as a function of time in one-dimensional model with random transmission
		amplitudes. Average transmission probability is taken to be $T=0.5$.
		(a)~Smaller randomness in absolute value transmission amplitude, $\delta T = 0.4$.
		(b)~Larger one, $\delta T = 0.98$.
		(c)~Smaller randomness in transmission amplitude phase,
		$\delta \chi_{\scriptscriptstyle\rm LL} = \delta \chi_{\scriptscriptstyle\rm LR} = 0.1\pi$.
		(d)~Larger one, $\delta \chi_{\scriptscriptstyle\rm LL} = \delta \chi_{\scriptscriptstyle\rm LR} =
		2\pi$.
	}
	\label{fig:1d_noiseAmp}
\end{figure}

The von Neumann entropy as a function of time is shown in Fig.~\ref{fig:1d_evolution}(d). While at some time steps the entropy may decrease, the long-time average entropy growth $S(\tau)\propto \log\tau$ is ensured by the entanglement with the spin reservoir having an infinite number of degrees of freedom. Numerical simulations shown in Fig.~\ref{fig:1d_log} confirms this logarithmic behavior. Position of the first entropy drop depends on transparency, e.g., it occurs between 2nd and 3rd steps at $T \in [0.35 \ldots 0.50]$, between 3rd and 4th steps at $T \in [0.58 \ldots 0.71]$, and between 4th and 5th steps at $T \in [0.72 \ldots 0.80]$.

\subsection{Randomness in transmission/reflection amplitudes
\label{sec:1d_noiseAmp}}

The entropy evolution for a model, where transmission and reflection amplitudes are different is shown in Fig.~\ref{fig:1d_noiseAmp}. We parametrize the scattering matrix in zero magnetic field by transparency $T$, phase of left reflection amplitude $\chi_{\scriptscriptstyle\rm LL}$ and phase of left-to-right transmission amplitude $\chi_{\scriptscriptstyle\rm LR}$. After that we randomize these parameters: $T_i = T + g_i \delta T$, $\chi_{{\scriptscriptstyle\rm LL},i} = \chi_{\scriptscriptstyle\rm LL} + g_i \delta \chi_{\scriptscriptstyle\rm LL}$, and $\chi_{{\scriptscriptstyle\rm LR},i} = \chi_{\scriptscriptstyle\rm LR} + g_i \delta \chi_{\scriptscriptstyle\rm LR}$, where~$g_i$ are homogeneously distributed in $[-0.5 \ldots 0.5]$.

\subsection{Randomness in the scatterer positions
\label{sec:1d_noise}}

\begin{figure}
	\includegraphics[width=7.5cm]{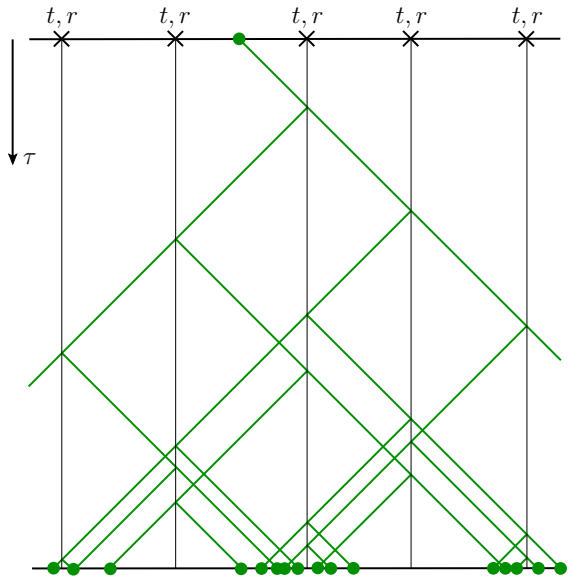}
	\caption{
		Illustration of the exponential growth of the number of components of
		the wave function with time in a system with the randomness in
		scatterers positions. We begin with a single infinitesimal wave packet,
		which moves to the right at $\tau=0$. Upon hitting the scatterer
		the wave function experiences partitioning, i.e.,
		each component splits into the two giving rise to the exponential growth
		of the number of components with time $\tau$.
		The mean distance between the components decreases.
	}
	\label{fig:1d_setup_noise_evolution}
\end{figure}

So far we were investigating regular arrays of scatterers/spins located at the integer positions ($x_i = i = 0$, $\pm 1$, $\pm 2, \ldots$). Now let us remove all spins and add random component to scatterer positions, $x_i = i + \eta g_i$, where $g_i$ are homogeneously distributed in $[-0.5 \ldots 0.5]$ and $\eta$ is a disorder strength in the system. In this case one needs to calculate all Feynman trajectories shown in Fig.~\ref{fig:1d_setup_noise_evolution} and density matrix. Naturally, now we cannot introduce discrete time and use continuous one. We start with a single trajectory, i.e., with the particle moving left or right with the amplitude equal to unity and the unity speed. The typical entropy evolution is shown in the inset in the Fig.~\ref{fig:1d_evolution}(d). This effect on entropy is very similar to what happens in the case of the random transmission/reflection amplitudes as in Sec.~\ref{sec:1d_noiseAmp}.

\begin{figure}[t]
	\includegraphics[width=8.5cm]{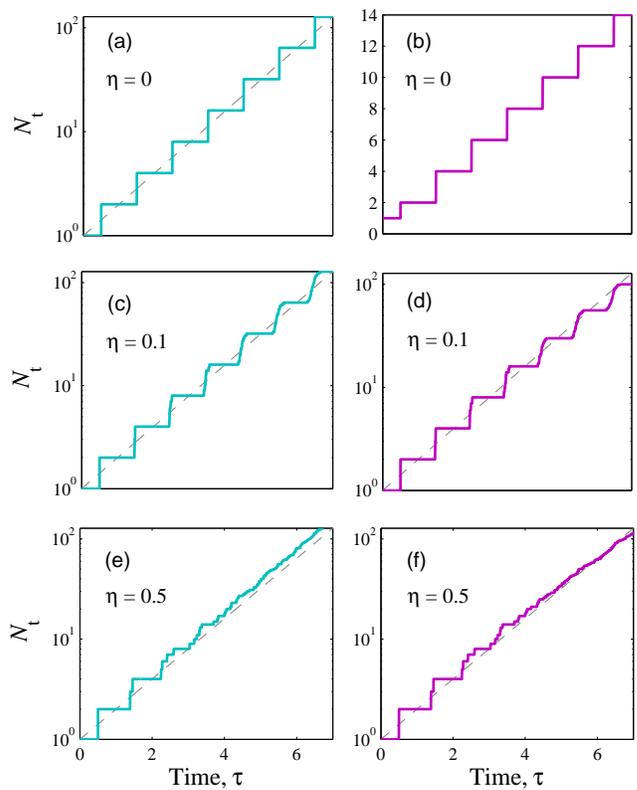}
	\caption{
		(a,b)~System with zero noise level $\eta=0$ behaves in a steplike
		manner and has $N_{\rm t} = 2^\tau = 1024$ trajectories and
		$N_{\rm wf} = 2\tau = 20$ wave function components at time
		$\tau=10$, see Sec.~\ref{sec:1d_equidistant}.
		(c,d)~In a system with small noise level $\eta=0.1$ steps smooth
		a little bit in time. But the number of trajectories rapidly increases, system
		has $N_{\rm t} = 1024$ trajectories and $N_{\rm wf} = 548$ components
		at $\tau=10$ (for the certain realization).
		(e,f)~In a noisy system $\eta=0.5$ steps smooth quickly.
		It has $N_{\rm t} = 1354$ trajectories and $N_{\rm wf} = 622$ components
		$\tau=10$.
	}
	\label{fig:1d_noise_trajectories}
\end{figure}

Let us consider several trajectories and the corresponding number of components of the wave function. Complexity of reversibility can be roughly estimated as number of wave function components $N_{\rm wf}$ at the time $\tau$; this gives approximately the number of the elementary operations needed for the reversal procedure. In case of the equidistant positions ($\eta = 0$), the number of trajectories is exponentially large $N_{\rm t} = 2^{\bar\tau}$, but due to the special interference conditions the number of wave function components is much smaller, $N_{\rm wf} = 2{\bar\tau}\ll N_{\rm t}$, where ${\bar\tau}$ is the closest discrete time to $\tau$. The step-like behavior of the number of trajectories is shown in Figs.~\ref{fig:1d_noise_trajectories}(a) and \ref{fig:1d_noise_trajectories}(b). Note that $N_{\rm wf}$ can be less if we start from special initial conditions, i.e., $N_{\rm wf} = {\bar\tau}$ if particle starts to move from the initial position in between two scatterers as described in Sec.~\ref{sec:1d_equidistant}. Upon adding small randomness ($\eta = 0.1$) the positions of the scatterers slightly shift. In the beginning of evolution all the scatterings and dephasing processes occur near the half-integer times and the amount of trajectories grow in a step-like manner. Further, the difference in trajectories accumulates and steps smoothen, see Figs.~\ref{fig:1d_noise_trajectories}(c) and \ref{fig:1d_noise_trajectories}(d). It is important to mention that the number of wave function components grows exponentially for any finite $\eta$. For larger randomness ($\eta = 0.5$) numbers $N_{\rm t}$ and $N_{\rm wf}$ quickly saturates at exponential behavior as shown in Figs.~\ref{fig:1d_noise_trajectories}(e) and \ref{fig:1d_noise_trajectories}(f).

\begin{figure}[t]
	\includegraphics[width=8.5cm]{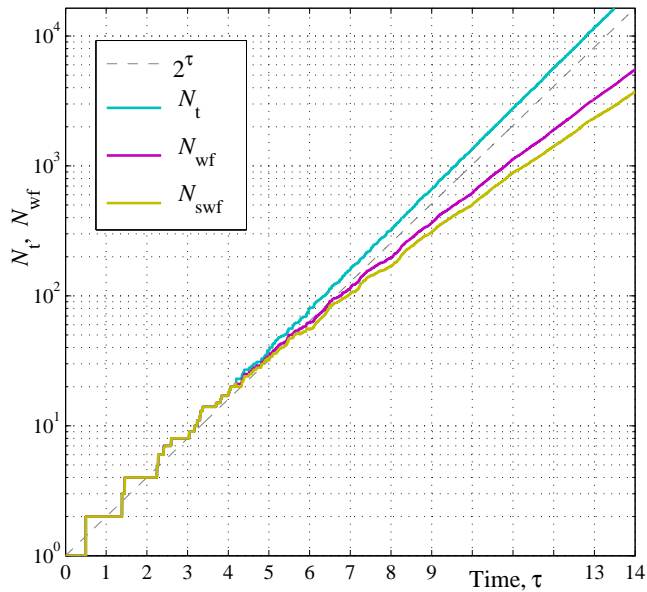}
	\caption{
		The number of trajectories $N_{\rm t}$ (cyan), the number of the wave function
		components $N_{\rm wf}$ (magenta), and the number of the significant wave
		function components comprising 99\% of the probability,
		$N_{\rm swf}$ (yellow) for an infinite 1D chain with the noise level, $\eta=0.5$.
		The number of trajectories grows faster then $2^\tau$ for noiseless 1D system.
		The number of the wave function components and the significant wave function
		components grows slightly sub-exponentially.
	}
	\label{fig:1d_noise_trajectories_fit}
\end{figure}

Shown in Fig.~\ref{fig:1d_noise_trajectories_fit} is the growth of $N_{\rm t}$, $N_{\rm wf}$, and the number of the significant wave function components $N_{\rm swf}$ comprising 99\% of probability density. We run the numerical simulations for $\eta = 0.5$ and wait while $N_{\rm t}$, $N_{\rm wf}$, and $N_{\rm swf}$ saturate on average. Fitting the curves one gets sub-exponential behavior for $N_{\rm wf} \approx 2^{1.25\tau^{0.87}}$ and $N_{\rm swf} \approx 2^{1.365\tau^{0.82}}$. This simple model demonstrates complexity of the reversibility in more realistic situations. Comparing features coming from the noise in transmission amplitudes (Sec.~\ref{sec:1d_noiseAmp}) and coming from noise in positions one can conclude that entropy is not always a good quantitative measure of reversibility complexity. The exponentially large number of wave function components implies that the reversal procedure would require an exponentially large number of manipulations to reverse the evolution of a quantum system subject to quenched disorder. Thus the effect of quenched disorder is similar to that of entanglement with the reservoir.

However, this exponential dependence saturates to a polynomial one as soon as the finite width of the wave packet [growing with time, see Eq.~(\ref{eq:wp_inf_t})] is taken into account, since the average distance between the zero-width wave packets decreases with time as shown Fig.~\ref{fig:1d_setup_noise_evolution}. Thus at some point the finite-width packets corresponding to the different components of the wave function start to overlap. As the wave packets mix up, the characteristic scale on which the variation of the wave function occurs can be estimated as $\lambda = 1/k_{\rm max}$, where $k_{\rm max}$ is the maximal $k$ in $f(k)$, see Eq.~(\ref{eq:wp_Fourier}). Then the number of the components of the wave function which will require the reversal procedure is about $L/\lambda$ in 1D and $(L/\lambda)^3$ in 3D, where $L$ is the size of the wave function. In the diffusion regime which realizes, in particular, above the localization threshold, $L \sim (D\tau)^{1/2}$. To conclude here, we have found that in the beginning of the evolution one observes the exponential growth of the number of components which after some time saturates into a polynomial one.


\begin{thebibliography}{99}

\bibitem{Boltzmann:1872}
	L.\,Boltzmann,
	Wiener Berichte {\bf 75}, 62--100 (1872).

\bibitem{Boltzmann:1896}
	L.\,Boltzmann,
	Annalen der Physik (Leipzig) {\bf 57}, 773--784 (1896)
	[Translated and reprinted in S.\,G.\,Brush, Kinetic Theory 2, Pergamon Elmsford, New York (1966)].

\bibitem{Lebowitz:1999}
	J.\,L.\,Lebowitz,
	Rev. Mod. Phys. {\bf 71}, S346--S357 (1999).

\bibitem{Boltzmann:1877}
	L.\,Boltzmann,
	Wiener Berichte {\bf 76}, 373--435 (1877).

\bibitem{Neumann:1929}
	J.\,von\,Neumann,
	Eur. Phys. J. H {\bf 35}, 201--237 (2010)
	[J. Von Neuman,
	Zeitschrift f\"ur Physik {\bf 57}, 30--70 (1929)].

\bibitem{Gemmer:2004}
	J.\,Gemmer, M.\,Micvhel, and G.\,Mahler,
	{\it Quantum thermodynamics}
	(Springer, Berlin, 2004).

\bibitem{Gemmer:2001}
	J.\,Gemmer, A.\,Otte, and G.\,Mahler,
	Phys. Rev. Lett. {\bf 86}, 1927--1930 (2001).

\bibitem{Popesku:2006}
	S.\,Popescu, A.\,J.\,Short, and A.\,Winter,
	Nat. Phys. {\bf 2}, 754--758 (2006).

\bibitem{Nielsen:2011}
	M.\,A.\,Nielsen and I.\,L.\,Chuang,
	{\it Quantum computation and quantum information}
	(Cambridge University Press, 2011).

\bibitem{Winter:2006}
	S.\,Popescu, A.\,J.\,Short, and A.\,Winter,
	Nat. Phys. {\bf 2}, 754--758 (2006).

\bibitem{Linden:2009}
	N.\,Linden, S.\,Popescu, A.\,J.\,Short, and A.\,Winter,
	Phys. Rev. E {\bf 79}, 061103 (2009).

\bibitem{Wehrl:1978}
	A.\,Wehrl,
	Rev. Mod. Phys. {\bf 50}, 221--260 (1978).

\bibitem{Holevo:2010}
	A.\,S.\,Holevo,
	Doklady Math. {\bf 82}, 730--731 (2010); arXiv:1003.5765.

\bibitem{Altshuler:1982}
	B.\,L.\,Altshuler, A.\,G.\,Aronov, D.\,E.\,Khmelnistkii, and A.\,I.\,Larkin,
	{\it Coherent effects in disordered conductors}
	in Quantum theory of solids, Ed. by I.\,M.\,Lifshits, 130--237 (Mir, 1982).

\bibitem{Burkard:2004}
	G.\,Burkard, R.\,H.\,Koch, and D.\,P.\,DiVincenzo,
	Phys. Rev. B {\bf 69}, 064503 (2004).

\bibitem{Lesovik:2013}
	G.\,B.\,Lesovik,
	Pis'ma v ZhETF {\bf 98}, 207--213 (2013).

\bibitem{Hahn:1950}
	E.\,L.\,Hahn,
	Phys. Rev. {\bf 80}, 580--594 (1950).

\bibitem{Viola:1999}
	L.\,Viola, E.\,Knill, and S.\,Lloyd,
	Phys. Rev. Lett. {\bf 82}, 2417--2421 (1999).

\bibitem{Poincare:1890}
	H.\,Poincar{\'e},
	Acta Mathematica {\bf 13}, 1--270 (1890).

\bibitem{Bocchieri:1957}
	P.\,Bocchieri and A.\,Loinger,
	Phys. Rev. {\bf 107}, 337--338 (1957).

\bibitem{Leith:1966}
	E.\,N.\,Leith and J.\,Upatneiks,
	J. Opt. Soc. Am. {\bf 56}, 523 (1966).

\bibitem{Boyd:2008}
	R.\,W.\,Boyd,
	{\it Nonlinear optics}
	(Academic Press, Amsterdam, 2008).
	
\bibitem{Andreev:2002}
	N.\,F.\,Andreev, V.\,I.\,Bespalov, M.\,A.\,Dvoretsky, and G.\,A.\,Pasmanik,
	IEEE J. Quantum Electron. {\bf 25}, 346--350 (2002).

\bibitem{Lebedev:2014}
	A.\,V.\,Lebedev, G.\,B.\,Lesovik, and G.\,Blatter, 
	to be published.

\bibitem{Haroshe:2008}
	S.\,Deleglise, I.\,Dotsenko, C.\,Sayrin, J.\,Bernu, M.\,Brune, J.-M.\,Raimond, and S.\,Haroche, 
	Nature {\bf 455}, 510--514 (2008).

\bibitem{Reed:1975}
	M.\,Reed and B.\,Simon,
	{\it Methods of modern mathematical physics II: Fourier analysis, self-adjointness}
	(Academic Press, New York, 1975).

\bibitem{Levitov:1996}
	L.\,S.\,Levitov, Lee\,H.\,W., and Lesovik\,G.\,B.,
	J. Math. Phys. {\bf 37}, 4845 (1996).

\end{thebibliography}
\end{document}